\documentclass[prl,twocolumn,showpacs]{revtex4}
\usepackage{amsmath}
\usepackage{epsfig}
\usepackage{amsfonts}
\begin{document}
\def\b{\bar}
\def\d{\partial}
\def\D{\Delta}
\def\cD{{\cal D}}
\def\cK{{\cal K}}
\def\f{\varphi}
\def\g{\gamma}
\def\G{\Gamma}
\def\l{\lambda}
\def\L{\Lambda}
\def\M{{\Cal M}}
\def\m{\mu}
\def\n{\nu}
\def\p{\psi}
\def\q{\b q}
\def\r{\rho}
\def\t{\tau}
\def\x{\phi}
\def\X{\~\xi}
\def\~{\widetilde}
\def\h{\eta}
\def\bZ{\bar Z}
\def\cY{\bar Y}
\def\bY3{\bar Y_{,3}}
\def\Y3{Y_{,3}}
\def\z{\zeta}
\def\Z{{\b\zeta}}
\def\Y{{\bar Y}}
\def\cZ{{\bar Z}}
\def\`{\dot}
\def\be{\begin{equation}}
\def\ee{\end{equation}}
\def\bea{\begin{eqnarray}}
\def\eea{\end{eqnarray}}
\def\half{\frac{1}{2}}
\def\fn{\footnote}
\def\bh{black hole \ }
\def\cL{{\cal L}}
\def\cH{{\cal H}}
\def\cF{{\cal F}}
\def\cP{{\cal P}}
\def\cM{{\cal M}}
\def\ik{ik}
\def\mn{{\mu\nu}}
\def\a{\alpha}

\title{Superconducting Source of the Kerr-Newman Electron}

\author{Alexander Burinskii}

\affiliation{Theoretical Physics Laboratory, NSI, Russian Academy of Sciences,\\
B. Tulskaya 52  Moscow 115191 Russia\footnote{E-mail address: bur@ibrae.ac.ru}}

\begin{abstract}
Regular superconducting  solution for interior of the Kerr-Newman
(KN) spinning particle is obtained. For parameters of electron it
represents a highly oblated rotating bubble formed by  Higgs field
which expels the electromagnetic (em) field and currents from interior
to domain wall boundary of the bubble.
 The external em and gravitational fields correspond exactly to
Kerr-Newman solution, while interior of the bubble is flat and
forms a `false' vacuum with zero energy density. Vortex of the KN
em field forms a quantum Wilson loop on the edge of the rotating
disk-like bubble.
\end{abstract}


\maketitle

1.{\it Introduction.} Kerr-Newman (KN) solution for a charged and
rotating Black-hole has $g=2$ as that of the Dirac electron and
paid attention as a classical model of electron coupled with
gravity
\cite{DKS,Isr,Bur0,Lop,BurSen,BurBag,RenGra,BurKN,Dym,AP,TN,KR}.
For parameters of electron (in the units $G=C=\hbar=1$) $J=1/2,
m\sim 10^{-22}, e^2\sim 137^{-1}$ the black-hole horizons
disappear and the Kerr singular ring of the Compton radius $a=
\hbar/2m \sim 10^{22}$ turns out to be naked. This ring forms the
gate to a negative sheet of the Kerr geometry, significance of
which is the old trouble for interpretation of the source of
Kerr geometry. The attempts by Brill and Cohen to match the
Kerr exterior with a rotating spherical shell and flat interior
\cite{BriCoh} have not lead to a consistent solution, and Israel
suggested to truncate the negative sheet, replacing it by a thin
(rotating) disk spanned by the Kerr ring \cite{Isr}. This model
led to a consistent source which was build from a very exotic
superluminal matter, besides the Kerr singular ring remained naked
at the edge of the disk. L\'opez \cite{Lop} transformed this model
to the model of an oblate rotating thin shell which covers the
Kerr ring. By especial choice of the boundary of the bubble,
$r=r_0=e^2/2m $ ($r$ is the Kerr oblate spheroidal coordinate), he
matched continuously the Kerr exterior with the flat interior, and
therefore he obtained a consistent KN source without the KN
singularity. However, like the other models, the L\'opez model was
not able to explain the origin of Poincar\'e stress, and a negative
pressure was introduces by L\'opez phenomenologically. Meanwhile,
the necessary tangential stress appears naturally in the domain walls
field models, in particular, in the models based on Higgs \cite{BurBag}.
It suggests that a consistent field descriptions of this problem
should contain, along with the KN Einstein-Maxwell sector, the
sector of Higgs fields and  sector of interaction between
the em field (coupled to gravity) and Higgs fields corresponding to
superconducting properties of the KN source. Up to our
knowledge, despite several attempts and partial results, no one
has been able so far to obtain a consistent field model of the KN source
related with Higgs field, and in this paper we present apparently first
solution of this sort which is consistent in the limit
of thin domain wall boundary of the bubble.
The considered KN source represents a generalization
of the L\'opez model and incorporate the results obtained earlier
in \cite{BEHM,BurBag,RenGra,BurCas}. Although we
consider here only the case of thin domain wall, generalization
to the case of a finite thickness is also possible by methods described
earlier in \cite{BurBag}. The described in  \cite{BurBag,BurCas}  KN
source was a bag formed by a potential  $V(r)$
interpolating between the external `true' vacuum $V^{(ext)}=0$ and a
`false' (superconducting) vacuum $V^{(in)}=0 $ inside of the
bag. It was shown that corresponding Higgs sector may be described
by $U(1)\times \tilde U(1)$ Witten
field model with the given by Morris in \cite{Mor} superpotential. It
was shown also in \cite{BurBag,BEHM,RenGra} that consistency of
the Einstein-Maxwell sector with such a phase transition may be
perfectly performed in the KS formalism with use of the  G\"urses
and G\"ursey ansatz \cite{GG}. However, the problem with consistency
appeared in \cite{BurBag} by the treatment of interaction between
the em and Higgs field. In this short note we improve this deficiency
which allows us to present consistent solution.
\medskip

2. {\it Phase transition in gravitational sector.} Following
\cite{BurBag,BEHM,RenGra}, for external region we  use the exact
KN solution in the Kerr-Schild (KS) form of metric
 \be g_\mn=\eta _\mn + 2 H k_\m k_\n \label{ksH} \ ,  \ee
where $\eta^\mn $ is metric of the auxiliary Minkowski background
in Cartesian coordinates $x^\m=(t,x,y,z).$ Electromagnetic (em) KN
field is given by vector potential \be A^\m_{KN} = Re \frac e
{r+ia \cos \theta} k^\m ,  \label{AKN} \ee where $k^\m(x^\m)$ is
the null vector field which is tangent to a vortex field of null
geodesic lines, the Kerr principal null congruence (PNC). For the
KN solution function $H$ has the form \be H=\frac {mr - e^2/2}{r^2
+ a^2 \cos ^2 \theta} \label{HKN}. \ee The used by Kerr especial
spherical oblate coordinates $r,\theta, \phi_K ,$ are related with
the Cartesian coordinates as follows

$ x+iy = (r + ia) e^{i\phi_K} \sin \theta , \  z=r\cos \theta .$
Vector field $k^\m$ is represented in the form \cite{DKS} \be k_\m
dx^\m = dr - dt - a \sin ^2 \theta d\phi_K . \label{km} \ee For
the metric {\it inside of the oblate bubble} we use the KS ansatz
(\ref{ksH}) in the form suggested by G\"urses and G\"ursey
\cite{GG} with function
\[ H=\frac {f(r)}{r^2 + a^2 \cos ^2 \theta} \ .\]
If we set for interior $f(r)=f_{int}=\alpha r^4 ,$ the Kerr singularity
will be suppressed.

 For exterior, $r>r_0, $ we use $f(r)=f_{KN}= mr -e^2/2$
corresponding to KN solution. Therefore, $f(r)$ describes a phase
transition of the KS metric from `true' to `false' vacuum (see
fig.1).

\begin{figure}[ht]
\centerline{\epsfig{figure=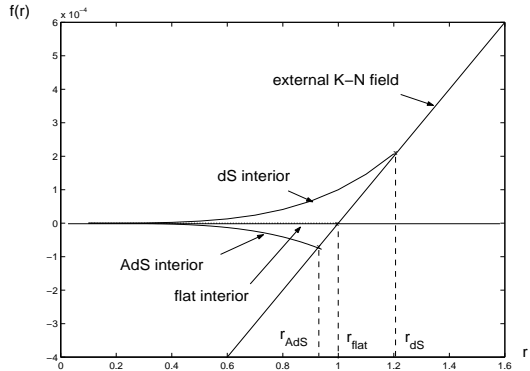,height=5cm,width=7cm}}
\caption{Matching of the metric for regular bubble interior with
metric of external KN field.}
\end{figure}

We assume that the zone of phase transition, $r\approx r_0
,$ is very thin and metric is continuous there, so
the point of intersection, $f_{int}(r_0)=f_{KN}(r_0),$ determines position of
domain wall $r_0$ graphically and yields the `balance matter
equation'\cite{RenGra,BEHM},
\be m= m_{em}(r_0)+ m_{mat}(r_0), \quad
m_{em}(r_0)=\frac {e^2} {2r_0} \label{bal-eq} ,\ee
which determines $r_0 $ analytically.
Interior has constant curvature, $\alpha=8\pi \Lambda/6 . $
For parameters of electron $a = J/m =1/2m >> r_0 = e^2/2m $
and the axis ratio of the ellipsoidal bubble is
$r_0/a = e^2\sim 137^{-1} ,$ so the bubble has the form of highly oblated disk.
\medskip

3. {\it Brief summary of the Higgs sector \cite{BurBag,BurCas}.}
The corresponding phase transition is provided by  Higgs model
with two complex Higgs field $\Phi$ and $\Sigma ,$ two related
gauge fields $A^\m$ and $B^\m ,$ and one auxiliary real field
$Z .$ This is a given by Morris \cite{Mor} generalization of the
$U(1)\times \tilde U(1)$ field model used by Witten for
superconducting strings \cite{Wit}.

The potential $V(r)=\sum _i |\d_i W|^2 ,$ where
$ \d_1 = \d_\Phi , \ \d_2 = \d_Z , \ \d_3 = \d_\Sigma ,$ is
determined by superpotential

$W=
\lambda Z(\Sigma \bar \Sigma -\eta^2) + (cZ+m) \Phi \bar \Phi ,$
where
$c, \ m, \ \eta, \ \lambda$ are real constants.
It forms a domain wall interpolating between the internal (`false')
and external (`true') vacua.

The vacuum states obey the conditions $\d_i W =0 $ which yield $V=0$

i) for `false'
vacuum ($r<r_0$): $Z=-m/c; \Sigma=0; |\Phi|=
\eta\sqrt{\lambda/c},$ as well as

ii) for `true' vacuum ($r>r_0$) : $
Z=0; \Phi=0; \Sigma=\eta .$
\medskip

4. {\it Interaction between the KN and Higgs fields.}
We set $B^\m =0 $ and use only
the gauge fields $A^\m $   which interacts with the Higgs field $
\Phi(x) = |\Phi(x)|e^{i \chi(x)}$ having a nonzero vev inside of
the bubble, $|\Phi(x)|_{r<r_0}=\Phi_0 ,$ and vanishing outside.
The related part of the field model is a copy of the
Nielsen-Olesen (NO) field model  for a vortex string in
superconducting media \cite{NO}, however, there is principal
difference in topology, since we describe superconducting interior
of the bubble, contrary to superconducting external media of the
NO model. The Lagrangian  ${\cal L}_{I}= -\frac 14 F_\mn F^\mn +
\frac 12 (\cD_\m \Phi)(\cD^\m \Phi)^* ,$ where $ \cD_\m =
\nabla_\m +ie A_\m ,$  leads to the equations $ \nabla _\n
\nabla^\n A_\m = I_\m = \frac 12 e |\Phi|^2 (\chi,_\m + e A_\m) .$
In external region $|\Phi |=0 $ and the em field has to correspond
to exact KN solution, $A^\m \equiv A^\m_{KN}.$ Following L\'opez,
we fix the boundary of bubble at $r_0=r_e=e^2/2m $ (one half of the
classical radius of electron), which yields flat interior,
$\alpha=0,$ and allows us to use flat d'Alemberian  and set
$\cD_\m = \d_\m +ie A_\m $ for $r<r_0 .$ We assume that current
has to be expelled from interior of the bubble to its boundary
(perfect superconductor), and set in interior $I_\m=0 ,$ which
yields \be \Box A_\m =0 = e |\Phi|^2 (\chi,_\m + e A^{(in)}_\m) .
\label{Main}\ee The KN gauge field $A_\m$ is given by (\ref{AKN})
and (\ref{km}). Inside of the bubble the specific Kerr angular coordinate
$\phi_K $  turns out to be inconsistent with the simple angular
coordinate of the Higgs field, $\phi=-i\ln[(x+iy)/\rho], \ \rho=(x^2+y^2)^{1/2} .$
Using the relation between corresponding differentials
$d\phi_K =\phi + \frac {adr}{r^2+a^2} ,$
we transform vector potential on the boundary and inside of the bubble
to the form \be A_\m dx^\m = \frac
{-er}{r^2+a^2\cos^2 \theta} [dt + a \sin ^2 \theta d\phi ] + \frac
{2e r dr} {(r^2 +a^2)} \label{Ain} \ee which shows that radial
component $A_r$ is full differential. At the external side of the bubble
$r=r_e +0 = e^2/2m ,$ the value of potential is $ A_\m dx^\m|_{r =
r_e +0} = \frac{-e r_e} {r^2_e +a^2 \cos^2 \theta}[dt + a \sin ^2
\theta d\phi ] + \frac {2e r_e dr} {(r_e^2 +a^2)} . $
The lines of KN vector-potential in equatorial plane  are tangent to the
Kerr singular ring, and  approaching the string-like edge of bubble by
$\cos\theta=0$ the
potential takes the form  \be A^{(str)}_\m dx^\m = - \frac{2m} {e}[dt +
a d\phi ] + \frac {2e r_e dr} {(r_e^2 +a^2)}. \ee
In particular, the tangent component at the edge is $A^{(str)}_\phi= -
2ma/e .$ Since $J=ma$  we have $A^{(str)}_\phi= - 2J/e .$
Setting $J=\frac n2 , n=1,2...$ we find out that vector potential forms on the edge
of bubble a closed quantized Wilson
loop \[S=\oint e
 A^{(str)}_\phi d\phi= - 2\pi n ,\] which has to be matched with angular
 periodicity of the Higgs field $\Phi = \Phi_0 \exp(i\chi )$ and fixes its $\phi$-dependence,  $\Phi \sim \exp \{i n \phi \} .$
For the time-like component of $A_\m$ inside of the bubble, the r.h.s. of (\ref{Main})
states $\chi,_0=-eA^{(in)}_0(r),$ which determines the $\chi,_0$ to be
a constant corresponding to frequency of
oscillations of the Higgs field, $\chi,_0=\omega=-eA^{(str)}_0=2m
.$ Radial component of the KN field is a full differential, and being extended
inside the bubble, it is compensated by Higgs field
in agreement with the r.h.s. of (\ref{Main}).
Therefore, the Higgs field acquires the form
 \be \Phi(x)= \Phi_0 \exp \{i\omega t - i \ln (r^2 +a^2) + i n\phi
  \} .\ee  For exclusion of the region of
 string-like loop at equator, the time and $\phi$ components of
 the gauge field have a chock crossing the boundary of bubble, which
 determine a distribution of circular currents over
 the bubble boundary.
 \medskip

 5. {\it Consistency.} We find out that inside of the bubble ${\cal D}_\m \Phi =
 i\Phi [\d_\m \chi + e A^{(in)}_\m] \equiv 0 .$ Together with the result that
 $V^{(in)}=0 ,$ it leads to vanishing of the stress-energy tensor of matter
 inside of the bubble,
$ T^{(int)}_{\mu \nu} = (D_{\mu} \Phi )\overline {(D_{\nu}\Phi )}
-\frac 12 g_{\mu \nu} [ (D_{\lambda} \Phi )\overline
{(D^{\lambda}\Phi )} ] ,$ and provides flatness of the interior in
agreement with our assumptions.

 Therefore, the obtained superconducting solution
 turns out to be consistent in the limit of the infinitely thin domain wall.
 It should be emphasized two important distinctions of this model from the
 typical soliton-like field models:

 i) oscillations
 of the Higgs field with frequency $\omega = 2m ,$ and
 ii) the appearance of quantum Wilson loop on the edge of KN disk.

  The model can also be generalized for the domain walls of a finite thickness $\delta ,$,
 if the parameter $r_0 $ is much greater then $\delta .$ In this case, as it was considered
 in \cite{BurBag}, one can use the flat domain wall approximation, and the
 exact field equations corresponding to finite region of phase transition may be
 integrated, at least numerically.
\medskip

{\it Acknowledgements.} Author thanks Theo Nieuwenhuizen for
hospitality in ITP  of Amsterdam University and discussions
stimulating reconsideration of this old problem.

\end{document}